# Epitaxial Growth and Band Structure of Te Film on Graphene

Xiaochun Huang,[1,2] Jiaqi Guan,[1,2] Bing Liu,[1,2] Shuya Xing,[1] Weihua Wang,[1]* and Jiandong Guo[1,2,3]*

[1]*Beijing National Laboratory for Condensed Matter Physics and Institute of Physics, Chinese Academy of Sciences, Beijing 100190, China*
[2]*Schoolf of Physical Sciences, University of Chinese Academy of Sciences, Beijing 100190, China*
[3]*Collaborative Innovation Center of Quantum Matter, Beijing 100871, China*

*Email: weihuawang@iphy.ac.cn and jdguo@ iphy.ac.cn

**Abstract**

Tellurium (Te) films with monolayer and few-layer thickness are obtained by molecular beam epitaxy on a graphene/6H-SiC(0001) substrate and investigated by *in situ* scanning tunneling microscopy and spectroscopy (STM/STS). We reveal that the Te films are composed of parallel-arranged helical Te chains flat-lying on the graphene surface, exposing the (1×1) facet of $(10\bar{1}0)$ of the bulk crystal. The band gap of Te films increases monotonically with decreasing thickness, reaching ~0.92 eV for the monolayer Te. An explicit band bending at the edge between the monolayer Te and graphene substrate is visualized. With the thickness controlled in atomic scale, Te films show potential applications of in electronics and optoelectronics.

The discovery of graphene has stimulated tremendous interests in the investigation of elementary two-dimensional (2D) materials, which usually have exotic physical and chemical properties distinct from their bulk counterparts.[1-3] Using molecular beam epitaxy (MBE), a host of elementary 2D materials, such as silicene,[4-7] germanene,[8-10] stanene,[11] borophene,[12,13] antimonene[14] and blue phosphorus,[15] are successfully fabricated. The observed extraordinary electronic structures demonstrated their promise in the applications of future nano- and opto-electronic devices.

The bulk single crystalline Te is a narrow band gap semiconductor (0.33 eV).[16] A recent theoretical work proposed three types of structure for monolayer Te, α-Te, β-Te and γ-Te, among which α-Te and β-Te are semiconducting with the band gap of 0.75 and 1.47 eV, respectively.[17] Therefore an important spectral range from mid-infrared (0.3 eV) to near-infrared (1.5 eV) can be achieved by controlling the thickness of Te films down to a single atomic layer.[18,19] On the other hand, Te atoms exhibit strong spin-orbit coupling due to their relatively high atomic number. Under external strains, topological phase transitions of Te single crystal have been predicted by theoretical studies.[20,21] Additionally, a plenty of novel quantum materials, *e.g.*, topological insulators $Bi/Sb_2Te_3$ and $ZrTe_5$, as identified by photoemission spectroscopy,[22-26] and topological semimetals $W/MoTe_2$,[27-30] all have Te atomic layer as the building block, implying the exotic properties of the monolayer Te. However, besides different types of Te nanoarchitectures achieved by various methods,[31-38] reports on epitaxial growth of Te films and characterizations at the atomic scale remain absent.

Here we report the van der Waals (vdW) epitaxy of Te films on the surface of graphene on 6H-SiC(0001) substrate by MBE. Monolayer and few-layer Te films are obtained. By the scanning tunneling microscopy (STM) investigations, we reveal that the Te films are composed of parallel-arranged helical Te chains flat-lying on the graphene surface, exposing the (1×1) facet of $(10\bar{1}0)$ of the bulk crystal. The band gap of Te films increases monotonically with decreasing thickness from 0.33 (bulk) to 0.92 eV (monolayer) as measured by scanning tunneling microscopy (STS), covering the spectral range from mid-infrared to near-infrared. Moreover, the spatially resolved STS spectra reveal the upward band bending corresponding to hole accumulation at the boundary of monolayer Te and graphene.

Experiments were performed in an ultrahigh-vacuum (base pressure $< 1.0 \times 10^{-10}$ Torr) MBE-STM combined system (Unisoku). Graphene was prepared on nitrogen-doped 6H-SiC(0001) substrate by thermal treatment at 1400 K to achieve the $(6\sqrt{3} \times 6\sqrt{3})$ reconstruction.[39,40] High-purity Te (99.999%) was evaporated from a standard Knudsen cell at 500 K. During the deposition, substrate temperature was kept at 300 K. After growth, the sample was transferred to the STM cryostat at 5 K, and a Pt-Ir tip was used for the STM/STS measurements.

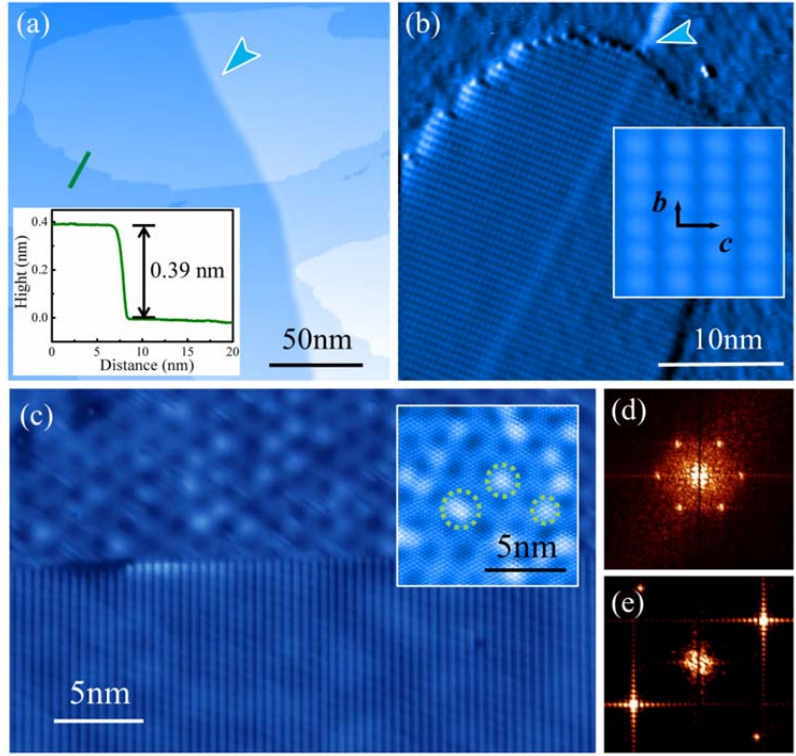

Figure 1. Topographic images of Te films on graphene. (a) Large area STM image (3.125 V / 50 pA) of the as-grown Te films spread across the substrate (SiC) step (indicated by blue arrow). The inset shows the line profile along the dark green line. (b) High-resolution STM image (1.5 V / 30 pA) of single-layer Te film. The rectangular lattice is labeled by the arrows *b* & *c* in the inset. (c) STM image (2.0 V / 30 pA) of the step between ($6\sqrt{3} \times 6\sqrt{3}$)-reconstructed graphene (upper part of the image) and the monolayer Te film (lower part of the image) with rectangular lattice. The inset shows the atomic-resolved STM image of graphene on SiC (-0.6 V / 50 pA) with the typical bright protrusions marked by green dashed circles. (d) and (e) Fast Fourier transform patterns of graphene and monolayer Te shown in (c), respectively.

Figure 1 shows the topography of Te films with different thickness on the graphene/SiC substrate after 10-minute deposition. Large-scale, atomically flat Te films are obtained, which can cover two adjacent terraces of the substrate continuously across the step (indicated by the arrow in Fig. 1a and b). By increasing deposition time, the films tend to expose enlarged terraces with increased thickness. The interspacing of Te layers (*d*) is measured at the step between adjacent Te terraces, and the profile along the dark green line gives $d = 3.9 \pm 0.1$ Å (inset in Figure 1a). Due to the different density of states, the measured step height between Te and graphene varies in the range of ~0.3 Å according to different tunneling parameters. The lowest step height between is measured as ~1.5 Å (Fig. 1b and c), indicating that monolayer Te films are obtained. In contrast to the hexagonal symmetry of $(6\sqrt{3} \times 6\sqrt{3})$-reconstructed graphene substrate, rectangular lattice is exhibited on Te terraces, with in-plane lattice constants of $b = 4.42 \pm 0.05$ Å and $c = 5.93 \pm 0.05$ Å. It is also visualized by the Fast Fourier transform (Figure 1d and e) of the STM image, signaling the vdW epitaxy of Te film on graphene. This is also evidenced by the arbitrary crystallographic orientations of Te films on the substrate.

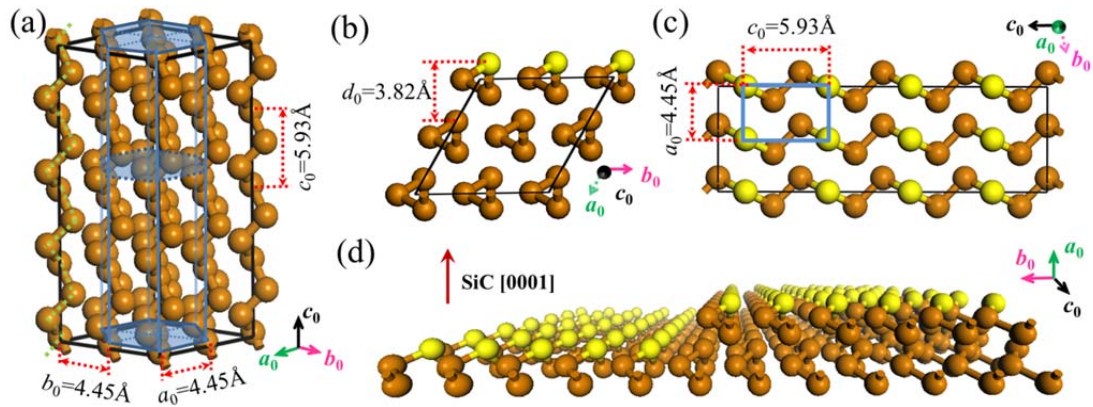

Figure 2. Lattice structure of Te single crystal with trigonal space group of $D_3^4$ (152#). (a) Schematic drawing of the 3D structure. The helical Te chain along $c_0$ axis is marked by the green dashed line. The blue hexagons indicate the centered hexagonal arrangement of Te chains. (b) and (c) The $a_0$-$b_0$ and $b_0$-$c_0$ planes, respectively. (d) 3D illustration of infinite bilayer Te films on graphene/SiC(0001), in which the dark red arrow indicated the SiC[0001] crystallographic orientation. The topmost Te atoms on each terrace are highlighted in (b), (c) and (d), and a rectangular primitive cell is marked by a blue rectangle.

To further elucidate the epitaxial relationship between Te films and graphene substrate, the structure of Te bulk crystal is analyzed in detail. Figure 2a presents the schematic of three-dimensional (3D) structure of Te crystal with trigonal space group of $D_3^4$.[21] The lattice constants are $a_0 = 4.45$ Å, $b_0 = 4.45$ Å and $c_0 = 5.93$ Å, respectively.[20] The Te crystal can be viewed as helical chains, in which Te atoms are covalently bonded with each other, arranged in an array of centered hexagons by weak vdW inter-chain interaction. On graphene with the in-plane lattice constant of 2.46 Å, the perpendicular crystallization of the helical chains is unfavorable due to the large lattice mismatch. Instead, the vdW epitaxial relationship leads the Te chains flat-lying on the substrate with the Te-Te covalent bonds saturated at the interface with the chemically inert graphene. As the consequence, the Te film composing of the helical chains arranged in parallel exposes the bulk $b_0$-$c_0$ (or $a_0$-$c_0$ that is equivalent) facet. Considering that one Te atom in every three along the helical chain sticks out to the topmost surface (highlighted in Fig. 2b and c), the rectangular lattice of Te films ($b = 4.42$ Å and $c = 5.93$ Å) observed by STM is quantitatively consistent with the bulk structure, *i.e.*, Te $(10\bar{1}0)$ film is grown on graphene with the bulk-truncated (1×1) structure. The observed step height of Te film (3.9 Å) also agrees with the interspacing of bulk $b_0$-$c_0$ faces (3.82 Å) in Figure 2b, confirming that Te films are stacked in the bulk arrangement (Figure 2d) and the terraces shown in Figure 1b and c correspond to monolayer Te films of parallel-arranged Te helical chains.

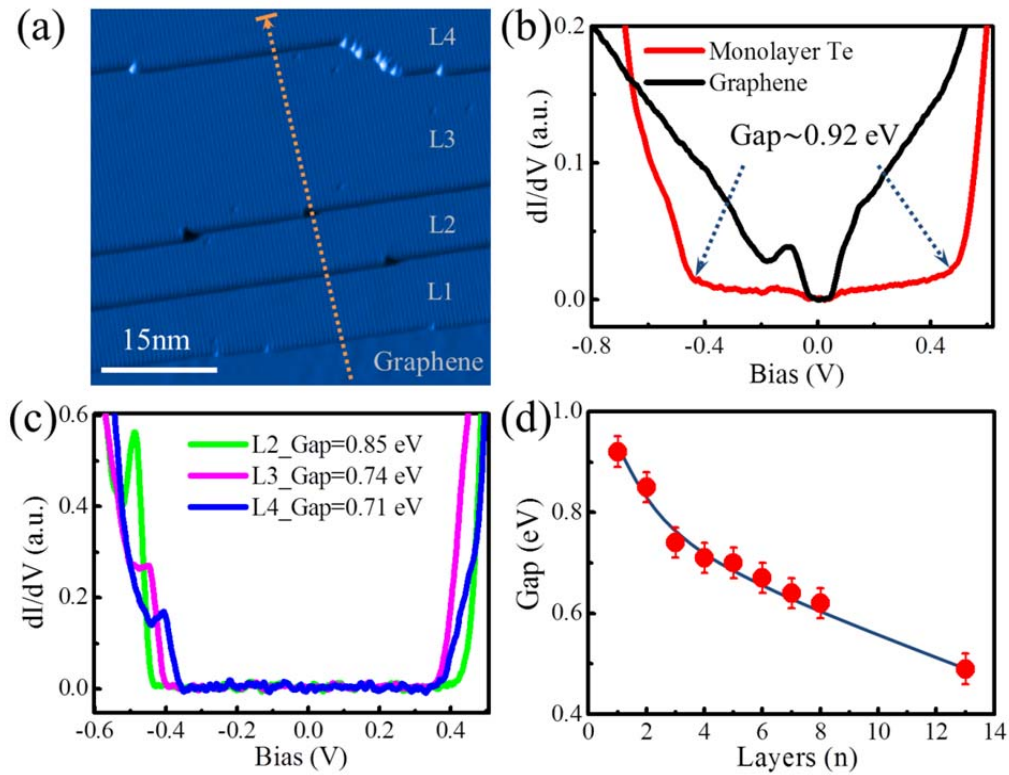

Figure 3. Band gap evolution of Te films with different thickness. (a) Differential STM image (-2 V / 20 pA) of a Te island on graphene/SiC exposing terraces of different layers, the first- (indicated by L1), second- (L2), third- (L3) and fourth- (L4). (b) dI/dV spectra (-800 mV / 100 pA) of both the graphene substrate and the monolayer Te. (c) dI/dV spectra taken on L2, L3 and L4, from which the band gap can be determined. All dI/dV spectra are averaged over 20 curves taken at the center of each terrace. (d) Band gap of Te films as a function of the thickness. The dark blue line is guide to the eye.

The vdW-type epitaxial (1×1) Te film on graphene/6H-SiC(0001) provides an ideal platform to investigate its intrinsic electronic structures. Additionally, some of the few-layer Te islands expose terraces corresponding to different thickness, as shown in Fig. 3a for example. This enables us to probe the thickness dependence of the band structure of Te films by direct comparison. As shown in Fig. 3b, the well-behaved d*I*/d*V* curve of graphene is consistent with that in Ref. 41, and a gap of 0.92 eV can be determined for the monolayer Te. Note that the finite density of states of graphene near the Fermi level influences the measurement on the monolayer Te, resulting the nonzero value of d*I*/d*V*. Explicit full band gaps with zero conductance inside are observed on the second- (0.85 eV), third- (0.74 eV) and fourth- (0.71 eV) layer of the Te films (Figure 3c). Figure 3d displays the thickness-dependent band gap up to 13 layers – it decreases monotonically with increasing thickness of Te films, from 0.92 (monolayer) to 0.49 eV (13-layer), towards the bulk gap of 0.33 eV.[16] The important mid- to near-infrared spectral range can be possibly covered by tuning the thickness of Te films with precision of single atomic layer.

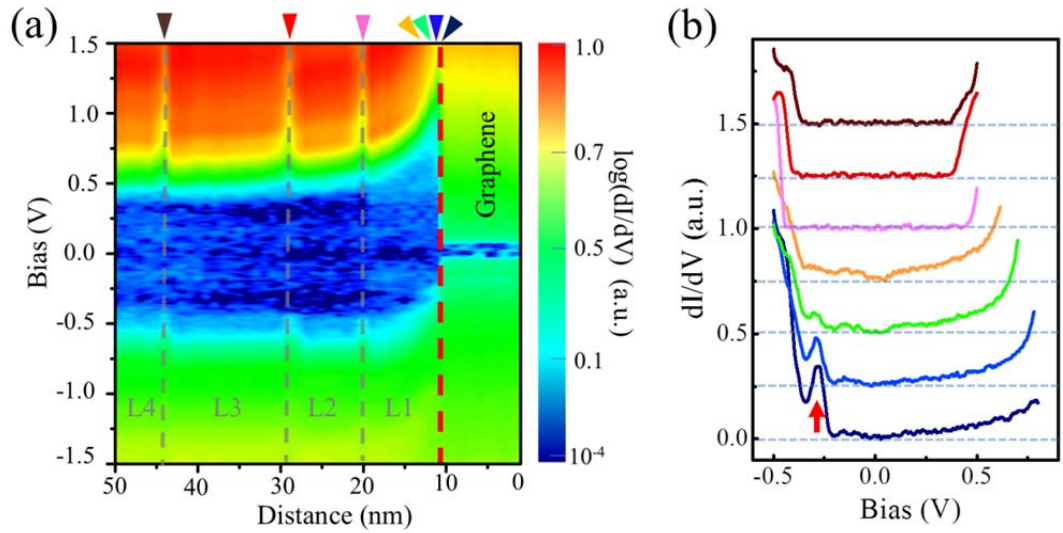

Figure 4. Differential conductance (dI/dV spectra) taken along the dashed orange line in Fig. 3(a). (a) The dI/dV spectra (50 curves taken with the interval of 1 nm) plotted in a 2D color mapping in logarithmic scale (-1.5 V / 100 pA, modulation of 10mV). The boundaries of monolayer Te-graphene and Te-Te are indicated by the red and gray dashed lines, respectively. (b) Representative dI/dV spectra taken at the positions labeled in (a) with the same color. The STS spectra are shifted along the perpendicular axis for clarity. The horizontal dashed lines indicate the zero conductance level for each spectrum.

STS maps along the dashed orange line in Figure 3a are performed to characterize the real space band profiles. As shown in Figure 4a, the band structure at the step edge between monolayer Te and graphene is distinct from other edges between Te terraces with different layers. An upward band bending of about 0.5 eV is observed at the boundary between monolayer Te and graphene, while it barely exists at the edges of Te-Te steps. This can be attributed to the Fermi level pinning by the in-gap states associated with the dangling bonds at the boundary between Te and graphene, where the Te lattice terminates and the interactions between helical chains is absent. Indeed an in-gap state at around -0.28 V (labeled by the red arrow in Figure 4b) is detected. It becomes suppressed remarkably and disappears rapidly away from the Te-graphene boundary. And it is not detected at all at Te-Te step edges. Similar Fermi level pining and the accompanying band bending have been observed at the boundary between $MoS_2$ and graphene.[42] As shown in Figure 4, the Te films with different thickness are nearly intrinsic semiconductor with the Fermi level locating in the middle of the band gap, while only at the Te-graphene boundary the upward band bending leads to the accumulation of holes.

It is worth mentioning that, although the vdW epitaxy has been considered the consequence of weak substrate effect, the Te-graphene interaction is manifested by the growth behaviors in the current work. The bright protrusions in STM image shown in the inset of Fig. 1c corresponding to the strain-induced wrinkles on graphene/SiC,[41,42] which may play an important role in the nucleation of Te growth. For comparison, we have grown Te film under the same condition on highly oriented pyrolytic graphite (HOPG) where such wrinkles are absent. No crystalline Te can be obtained. The Te-graphene interaction may also trigger controllable novel electronic and optoelectronic phenomena.

In summary, we grow monolayer and few-layer Te films on the graphene/6H-SiC(0001) substrate via vdW epitaxy. Te films with different thickness are formed in the structure of bulk Te crystal along $[10\bar{1}0]$ direction, with the helical chains flat-lying in each layer and being parallel to each other. The band gap of Te increases monotonically with decreasing thickness, up to 0.92 eV for the monolayer Te. Besides, an upward band bending is observed at the boundary of monolayer Te and graphene, manifesting the local hole accumulation due to interface interactions. This implies the possibility of tuning the electronic and optoelectronic properties of 2D Te films.

## AUTHOR INFORMATION

### Corresponding Author

Weihua Wang (weihuawang@iphy.ac.cn) and Jiandong Guo (jdguo@iphy.ac.cn)

### Notes

The authors declare no competing financial interest.

## ACKNOWLEDGMENTS

We thank Dr. Xv Wu and Zhilin Li for helpful discussions. This work is supported by the National Key Research and Development Program of China (2016YFA0300600, 2016YFA0202300), Chinese NSF (11634016 & 11474334) and the "Strategic Priority Research Program (B)" of the Chinese Academy of Sciences (XDB07030100). WHW is grateful to the financial support of the Hundred Talents Program of the Chinese Academy of Sciences.

## REFERENCES

[1]   Novoselov, K. S.; Geim, A. K.; Morozov, S. V.; Jiang, D.; Katsnelson, M. I.; Grigorieva, I. V.; Dubonos, S. V.; Firsov, A. A. *Nature* **2005,** 438, (7065), 197-200.

[2]   Geim, A. K.; Novoselov, K. S. *Nat. Mater.* **2007**, *6* (3), 183.

[3]   Castro Neto, A. H.; Guinea, F.; Peres, N. M. R.; Novoselov, K. S.; Geim, A. K. *Rev. Mod. Phys.* **2009,** 81, (1), 109-162.

[4]   Vogt, P.; De Padova, P.; Quaresima, C.; Avila, J.; Frantzeskakis, E.; Asensio, M. C.; Resta, A.; Ealet, B.; Le Lay, G. *Phys. Rev. Lett.***2012,** 108, (15), 155501.

[5]   Feng, B.; Ding, Z.; Meng, S.; Yao, Y.; He, X.; Cheng, P.; Chen, L.; Wu, K. *Nano Lett.* **2012,** 12, (7), 3507-11.

[6]   Fleurence, A.; Friedlein, R.; Ozaki, T.; Kawai, H.; Wang, Y.; Yamada-Takamura, Y. *Phys. Rev. Lett.* **2012,** 108, (24), 245501.

[7]   De Crescenzi, M.; Berbezier, I.; Scarselli, M.; Castrucci, P.; Abbarchi, M.; Ronda, A.; Jardali, F.; Park, J.; Vach, H. *ACS Nano* **2016,** 10, (12), 11163-11171.

[8]   Li, L.; Lu, S. Z.; Pan, J.; Qin, Z.; Wang, Y. Q.; Wang, Y.; Cao, G. Y.; Du, S.; Gao, H. J. *Adv. Mater.* **2014,** 26, (28), 4820-4.

[9]   Derivaz, M.; Dentel, D.; Stephan, R.; Hanf, M. C.; Mehdaoui, A.; Sonnet, P.; Pirri, C. *Nano Lett.* **2015,** 15, (4), 2510-6.

[10]  Zhang, L.; Bampoulis, P.; Rudenko, A. N.; Yao, Q.; van Houselt, A.; Poelsema, B.; Katsnelson, M. I.; Zandvliet, H. J. *Phys. Rev. Lett.* **2016,** 116, (25), 256804.

[11]  Zhu, F. F.; Chen, W. J.; Xu, Y.; Gao, C. L.; Guan, D. D.; Liu, C. H.; Qian, D.; Zhang, S. C.; Jia, J. F. *Nat. Mater.* **2015,** 14, (10), 1020-5.

[12] Mannix, A. J.; Zhou, X.-F.; Kiraly, B.; Wood, J. D.; Alducin, D.; Myers, B. D.; Liu, X.; Fisher, B. L.; Santiago, U.; Guest, J. R.; Yacaman, M. J.; Ponce, A.; Oganov, A. R.; Hersam, M. C.; Guisinger, N. P. *Science* **2015,** 350, (6267), 1513-1516.

[13] Feng, B.; Zhang, J.; Zhong, Q.; Li, W.; Li, S.; Li, H.; Cheng, P.; Meng, S.; Chen, L.; Wu, K. *Nat. Chem.* **2016,** 8, (6), 563-8.

[14] Wu, X.; Shao, Y.; Liu, H.; Feng, Z.; Wang, Y.-L.; Sun, J.-T.; Liu, C.; Wang, J.-O.; Liu, Z.-L.; Zhu, S.-Y.; Wang, Y.-Q.; Du, S.-X.; Shi, Y.-G.; Ibrahim, K.; Gao, H.-J. *Adv. Mater.* DOI: 10.1002/adma.201605407.

[15] Zhang, J. L.; Zhao, S.; Han, C.; Wang, Z.; Zhong, S.; Sun, S.; Guo, R.; Zhou, X.; Gu, C. D.; Yuan, K. D.; Li, Z.; Chen, W. *Nano Lett.* **2016,** 16, (8), 4903-8.

[16] Anzin, V. B.; Eremets, M. I.; Kosichkin, Y. V.; Nadezhdinskii, A. I.; Shirokov, A. M. *Phys. Status Solidi (a)* **1977,** 42, (1), 385-390.

[17] Zhu, L. Z.; Cai, X. L.; Niu, C. Y.; Yi, S. H.; Guo, Z. X; Liu, F.; Cho, J. H.; Jia, Y.; Zhang, Z. Y.; **2017,** *arXiv:1701.08875*.

[18] Li, L.; Kim, J.; Jin, C.; Ye, G. J.; Qiu, D. Y.; da Jornada, F. H.; Shi, Z.; Chen, L.; Zhang, Z.; Yang, F.; Watanabe, K.; Taniguchi, T.; Ren, W.; Louie, S. G.; Chen, X. H.; Zhang, Y.; Wang, F. *Nat. Nanotechnol.* **2017,** 12, (1), 21-25.

[19] Churchill, H. O.; Jarillo-Herrero, P. *Nat. Nanotechnol.* **2014,** 9, (5), 330-1.

[20] Agapito, L. A.; Kioussis, N.; Goddard, W. A., III; Ong, N. P. *Phys. Rev. Lett.* **2013,** 110, (17), 176401.

[21] Hirayama, M.; Okugawa, R.; Ishibashi, S.; Murakami, S.; Miyake, T. *Phys. Rev.*

*Lett.* **2015,** 114, (20), 206401.

[22] Hsieh, D.; Xia, Y.; Qian, D.; Wray, L.; Meier, F.; Dil, J. H.; Osterwalder, J.; Patthey, L.; Fedorov, A. V.; Lin, H.; Bansil, A.; Grauer, D.; Hor, Y. S.; Cava, R. J.; Hasan, M. Z. *Phys. Rev. Lett.* **2009,** 103, (14), 146401.

[23] Chen, X.; Ma, X. C.; He, K.; Jia, J. F.; Xue, Q. K. *Adv. Mater.* **2011,** 23, (9), 1162-5.

[24] Wang, G.; Zhu, X.-G.; Sun, Y.-Y.; Li, Y.-Y.; Zhang, T.; Wen, J.; Chen, X.; He, K.; Wang, L.-L.; Ma, X.-C.; Jia, J.-F.; Zhang, S. B.; Xue, Q.-K. *Adv. Mater.* **2011,** 23, (26), 2929-2932.

[25] Zhang, Y.; Wang, C. L.; Yu, L.; Liu, G. D.; Liang, A. J.; Huang, J. W.; Nie, S. M.; Zhang, Y. X.; Shen, B.; Liu, J.; Weng, H. M.; Zhao, L. X.; Chen, G. F.; Jia, X. W.; Hu, C.; Ding, Y.; He, S. L.; Zhao, L.; Zhang, F. F.; Zhang, S. J.; Yang, F.; Wang, Z. M.; Peng, Q. J.; Dai, X.; Fang, Z.; Xu, Z. Y.; Chen, C. T.; Zhou, X. J. **2016**, *arXiv:1602.03576*.

[26] Weng, H.; Dai, X.; Fang, Z. *Phys. Rev. X* **2014,** 4, (1) 011002.

[27] Wang, C.; Zhang, Y.; Huang, J.; Nie, S.; Liu, G.; Liang, A.; Zhang, Y.; Shen, B.; Liu, J.; Hu, C.; Ding, Y.; Liu, D.; Hu, Y.; He, S.; Zhao, L.; Yu, L.; Hu, J.; Wei, J.; Mao, Z.; Shi, Y.; Jia, X.; Zhang, F.; Zhang, S.; Yang, F.; Wang, Z.; Peng, Q.; Weng, H.; Dai, X.; Fang, Z.; Xu, Z.; Chen, C.; Zhou, X. J. *Phys. Rev. B* **2016,** 94, (24) 241119.

[28] Wu, Y.; Mou, D.; Jo, N. H.; Sun, K.; Huang, L.; Bud'ko, S. L.; Canfield, P. C.; Kaminski, A. *Phys. Rev. B* **2016,** 94, (12). 121113.

[29] Deng, K.; Wan, G.; Deng, P.; Zhang, K.; Ding, S.; Wang, E.; Yan, M.; Huang, H.; Zhang, H.; Xu, Z.; Denlinger, J.; Fedorov, A.; Yang, H.; Duan, W.; Yao, H.; Wu, Y.; Fan, S.; Zhang, H.; Chen, X.; Zhou, S. *Nat. Phys.* **2016,** 12, (12), 1105-1110.

[30] Jiang, J.; Liu, Z. K.; Sun, Y.; Yang, H. F.; Rajamathi, R.; Qi, Y. P.; Yang, L. X.; Chen, C.; Peng, H.; Hwang, C. C.; Sun, S. Z.; Mo, S. K.; Vobornik, I.; Fujii, J.; Parkin, S. S. P.; Felser, C.; Yan, B. H.; Chen, Y. L. **2016**, *arXiv:1604.00139*.

[31] Wang, Q.; Safdar, M.; Xu, K.; Mirza, M.; Wang, Z.; He, J. *ACS Nano* **2014,** 8, (7), 7497-7505.

[32] Xia, Y.; Yang, P.; Sun, Y.; Wu, Y.; Mayers, B.; Gates, B.; Yin, Y.; Kim, F.; Yan, H. *Adv. Mater.* **2003,** 15, (5), 353-389.

[33] Mayers, B.; Xia, Y. *Advanced materials* **2002,** 14, (4), 279-282.

[34] Mo, M.; Zeng, J.; Liu, X.; Yu, W.; Zhang, S.; Qian, Y. *Adv. Mater.* **2002,** 14, (22), 1658-1662.

[35] Zhu, Y.-J.; Wang, W.-W.; Qi, R.-J.; Hu, X.-L. *Angew. Chem.* **2004,** 116, (11), 1434-1438.

[36] Hawley, C. J.; Beatty, B. R.; Chen, G.; Spanier, J. E. *Cryst. Growth & Des.* **2012,** 12, (6), 2789-2793.

[37] Safdar, M.; Zhan, X.; Niu, M.; Mirza, M.; Zhao, Q.; Wang, Z.; Zhang, J.; Sun, L.; He, J. *Nanotechnol.* **2013,** 24, (18), 185705.

[38] Tai, G. a.; Zhou, B.; Guo, W. *The J. of Phys. Chem. C* **2008,** 112, (30), 11314-11318.

[39] Hass, J.; de Heer, W. A.; Conrad, E. H. *J. of Phys.: Condens. Matter* **2008,** 20,

(32), 323202.

[40] Wang, Q.; Zhang, W.; Wang, L.; He, K.; Ma, X.; Xue, Q. *J. Phys.: Condens. Matter* **2013,** 25, (9), 095002.

[41] Lauffer, P.; Emtsev, K. V.; Graupner, R.; Seyller, T.; Ley, L.; Reshanov, S. A.; Weber, H. B. *Phys. Rev. B* **2008,** 77, (15), 155426.

[42] Zhang, C.; Johnson, A.; Hsu, C. L.; Li, L. J.; Shih, C. K. *Nano Lett.* **2014,** 14, (5), 2443-7.

[43] Mallet, P.; Varchon, F.; Naud, C.; Magaud, L.; Berger, C.; Veuillen, J. Y. *Phys. Rev. B* **2007,** 76, (4), 041403.